\documentclass[aps,prl,twocolumn]{revtex4}

 \def\be{\begin{eqnarray}}
\def\ee{\end{eqnarray}}

\usepackage{graphicx}

\begin{document}

\title{Failure of Mean Field Theory at Large N}
\author{Shailesh Chandrasekharan}
\affiliation{
Department of Physics, Box 90305, Duke University,
Durham, North Carolina 27708.}
\author{Costas G. Strouthos}
\affiliation{
Department of Physics, University of Cyprus, CY-1678 Nicosia, Cyprus.}


\begin{abstract}
We study strongly coupled lattice QCD with $N$ colors of staggered
fermions in $3+1$ dimensions. While mean field theory describes the 
low temperature behavior of this theory at large $N$, it fails
in the scaling region close to the finite temperature second order 
chiral phase transition. The universal critical region close to
the phase transition belongs to the 3d $XY$ 
universality class even when $N$ becomes large. This is in contrast 
to Gross-Neveu models where the critical region shrinks as $N$ 
(the number of flavors) increases and mean field theory is 
expected to describe the phase transition exactly in the limit of 
infinite $N$. Our work demonstrates that close to second order 
phase transitions infrared fluctuations can sometimes be important
even when $N$ is strictly infinite.
\end{abstract}

\maketitle

Mean field techniques provide a simple but powerful approach to
gain qualitative insight of the underlying physics in a variety of 
field theories \cite{Zin}. The Bardeen-Cooper-Schrieffer (BCS) solution
to superconductivity is a well known application of such a technique.
Wilson's renormalization group shows that when correlation lengths 
$\xi$, associated with the fluctuations of the field, become 
large compared to the microscopic length scale $a$ of the problem, 
mean field techniques become exact in dimensions greater than four. 
However, in lower dimensional systems infrared fluctuations can 
become important when $\xi \gg a$ and invalidate the mean field 
arguments. For this reason the mean field approach must be used 
with care close to second order phase transitions. The region 
close to the critical point where mean field theory fails is 
usually referred to as the Ginsburg region \cite{Amit}. Using 
field theoretic techniques sometimes it is possible to estimate the 
Ginsburg region. In conventional superconductors the Ginsburg region 
is known to be suppressed by some power of $T_c/E_f$ where $T_c$ is 
the critical temperature and $E_f$ is the Fermi energy. Mean field
theory is believed to be reliable outside the Ginsburg region.
This is the reason why BCS is a good approach to understand the 
physics of superconductivity for all temperatures except very close 
to $T_c$.

There are certain limits in which the infrared fluctuations can 
be naturally suppressed even in low dimensional systems. For 
example consider a theory containing a field with $N$ components. 
When $N$ is large a saddle point approximation can be used to solve 
the theory and the leading term is nothing but the mean field solution. 
Theoretical physicists often use this feature to solve a physical 
theory by increasing the number of components of the field artificially.
By solving the theory in the limit of large number of components and 
computing the leading corrections they can sometimes estimate realistic 
answers in the physical theory. A variety of field theories can be 
studied in the large $N$ limit \cite{Mos03}. 

Although large $N$ approach to field theories has been very
successful, in this article we show that not all large $N$ limits 
lead to the suppression of infrared fluctuations. We contrast two 
models, the Gross-Neveu (GN) model involving $N$ species of 
fermions studied recently \cite{Koc95,Kog99} and strongly coupled 
$U(N)$ lattice QCD with staggered fermions (SCLQCD) studied here. 
Both these models contain a global symmetry which is spontaneously 
broken at low temperatures. Further, the symmetry group and the 
breaking pattern is not affected by $N$. These models undergo a 
second order phase transition to a symmetric phase at a finite 
critical temperature $T_c$. In the large $N$ limit they can be solved 
exactly using mean field techniques at low temperatures. An interesting 
question then is whether the mean field description is valid even 
close to $T_c$ when $N$ becomes infinite.

It was discovered in \cite{Koc95} that indeed in the GN model the 
critical behavior near $T_c$ belongs to the Landau-Ginsburg mean 
field universality class at large $N$. This implies that the Ginsburg 
region, where the critical behavior is governed by a non-trivial 
universality class that depends on the symmetry group and the breaking 
pattern in a dimensionally reduced theory, has a zero width at large $N$.
It was later shown in \cite{Kog99} 
that indeed the Ginsburg region is suppressed by a factor $1/\sqrt{N}$.
Although a $Z_2$ symmetric GN model was analyzed in \cite{Koc95,Kog99}, 
there are reasons to believe that the arguments would hold for 
continuous symmetries as well \cite{Ste04,Cos04}.

As we will demonstrate here, in contrast to the GN model, the Ginsburg 
region does not shrink in SCLQCD in the large $N$ limit. While mean field 
theory is indeed a good approximation at low temperatures, the finite 
temperature phase transition is not described by mean field theory even at 
infinite $N$.
We believe our results should be of interest to a wide range of physicists
since our model can be mapped into a theory of classical dimers. Dimer
models have a long history \cite{Hei72}. In the 1960s these models 
attracted a lot of attention  when it was shown that the Ising model can 
be rewritten as a dimer model \cite{Kas63,Fis63}. In the  late 1980s they 
gained popularity again in their quantum version \cite{Rok88} 
as a promising approach to the famous resonating-valence-bond (RVB) liquid 
phase \cite{And87}. More recently, this approach has gained momentum 
again since it was shown that the RVB phase was indeed realized on a 
triangular lattice but not on a cubic lattice \cite{Moe01,Moe02}. Thus 
physicists attempting to use large $N$ techniques in dimer models can
benefit from our results.

The partition function of SCLQCD is given by
\begin{equation}
Z(T,m) = \int [dU] [d\psi d\bar\psi]\ \exp\left(-S[U,\psi,\bar\psi]\right),
\label{uNpf}
\end{equation}
where $[dU]$ is the Haar measure over $U(N)$ matrices and $[d\psi d\bar\psi]$ 
specify Grassmann integration. The Euclidean space action 
$S[U,\psi,\bar\psi]$ in the strong (gauge) coupling limit with staggered
fermions is given by
\begin{equation}
\label{fact}
- \sum_{x,\mu} \frac{\eta_{x,\mu}}{2}\Big[\bar\psi_x U_{x,\mu} 
\psi_{x+\hat{\mu}}
- \bar\psi_{x+\hat{\mu}} U^\dagger_{x,\mu} \psi_x\Big]
- m \sum_x \bar\psi_x\psi_x,
\end{equation}
where $x$ refers to the lattice site on a periodic four-dimensional 
hyper-cubic lattice of size $L$ along the three spatial directions 
and size $L_t$ along the euclidean time direction. The index 
$\mu=1,2,3,4$ refers to the four space-time directions, $U_{x,\mu} \in U(N)$ 
is the usual links matrix representing the gauge fields, and 
$\psi_x,\bar\psi_x$ are the three-component staggered quark fields. The 
gauge fields satisfy periodic boundary conditions while the quark fields 
satisfy either periodic or anti-periodic boundary conditions. The factors 
$\eta_{x,\mu}$ are the well-known staggered fermion phase factors. Using 
an asymmetry factor between space and time we introduce a temperature in 
the theory. We choose $\eta_{x,\mu}^2 = 1, \mu=1,2,3$ (spatial 
directions) and $\eta_{x,4}^2 = T$ (temporal direction), where the real 
parameter $T$ is a coupling that controls the temperature. By working 
on anisotropic lattices with $L_t << L$ at fixed $L_t$ and varying $T$ 
continuously one can study finite temperature phase transitions 
\cite{Boy92}. In this article we fix $L_t=4$ for convenience.

The partition function given in eq.(\ref{uNpf}) can be rewritten as a
partition function for a monomer-dimer system, which is given by
\begin{equation}
Z(T,m) \;=\; \sum_{[n,b]} \;
\prod_{x,\mu}\; (z_{x,\mu})^{b_{x,\mu}}\frac{(N-b_{x,\mu})!}{b_{x,\mu}! N!}
\; \prod_x \frac{N!}{n_x!}\;m^{n_x},
\label{pf}
\end{equation}
and is discussed in detail in \cite{Ros84,Cha03}. Here $n_x=0,1,2,..,N$ 
refers to the number of monomers on the site $x$,  $b_{x,\mu}=0,1,2,...,N$ 
represents the number of dimers on the bond connecting $x$ and $x+\hat{\mu}$, 
$m$ is the monomer weight, $z_{x,\mu}=\eta_{x,\mu}^2/4$ are the dimer 
weights. Note that while spatial dimers carry a weight $1/4$, temporal 
dimers carry a weight $T/4$. The sum is over all monomer-dimer 
configurations $[n,b]$ which are constrained such that at each site,
$n_x + \sum_\mu [b_{x,\mu} + b_{x-\hat{\mu},\mu}] = N$.

When $m=0$, the action of SCLQCD, eq. (\ref{fact}), is invariant under 
$O(2)$ chiral transformations: 
$\psi_x \rightarrow \exp(i\sigma_x \theta)\psi_x$ and 
$\bar\psi_x \rightarrow \bar\psi_x \exp(i\sigma_x \theta)$ where 
$\sigma_x = 1$ for all even sites and $-1$ for all odd sites.
In the large $N$ limit mean field techniques can be used to show that this
chiral symmetry is spontaneously broken at low temperatures \cite{Klu81}. 
In \cite{Ros84} a Monte Carlo method was developed to solve the problem 
from first principles and it was shown that mean field theory is indeed 
reliable at small temperatures \cite{Ros84}. Unfortunately, since the 
algorithm was inefficient at small quark masses, the finite temperature 
chiral phase transition was never studied in the large $N$ limit. 
Recently a very efficient cluster algorithm was discovered to 
solve the model at any value of $N$ \cite{Ada03}. Using this algorithm it 
was shown with great precision that for $N=3$ the finite temperature phase 
transition belonged to the 3d $XY$ universality class \cite{Cha03}. Here we 
extend that calculation to higher values of $N$.

The order parameter that signals chiral symmetry breaking is the chiral 
condensate, defined by
\begin{equation}
\Sigma = \lim_{m\rightarrow 0}\lim_{L\rightarrow \infty}
\ \ 
\frac{1}{N L_t L^3} \frac{1}{Z}\frac{\partial}{\partial m}  Z(T,m).
\end{equation}
For a fixed $T$ the large $N$ result for $\Sigma$ can easily be obtained 
by extending the calculation of \cite{Ros84}. One gets
\begin{equation}
\Sigma = 
\sqrt{\frac{-9-17 T + 18 \sqrt{9 - T + T^2}}{81 - 18 T + T^2}}
\label{largen}
\end{equation}
which shows that the critical temperature, as we have defined it, is 
infinite. A calculation which includes the $1/N$ correction shows that 
$T_c \sim N$. In the large $d$ (spatial dimensions) limit one obtains
$T_c = d(N+2) L_t/6$ \cite{Bil92}.

In order to determine $\Sigma$ using Monte Carlo calculations we measure 
the chiral susceptibility in the chiral limit,
\begin{equation}
\chi = \lim_{m\rightarrow 0} 
\frac{1}{L^3} \frac{1}{Z}\frac{\partial^2}{\partial m^2} Z(T,m),
\end{equation}
The finite size scaling of this quantity is known from chiral perturbation
theory \cite{Has90} and one expects
\begin{equation}
\chi = \frac{N^2 L_t^2 \Sigma^2 L^3}{2}\Big[ 1 + \frac{0.226}{F^2 L} + 
\frac{\alpha}{L^2} +...\Big].
\label{chptchi}
\end{equation}
The constant $F^2$ is equal to
\begin{equation}
\lim_{L\rightarrow \infty}
\frac{1}{3 L^3}\Bigg\langle 
\Big\{[\sum_x J_{x,1}]^2 + [\sum_x J_{x,2}]^2 + [\sum_x J_{x,3}]^2 \Big\}
\Bigg\rangle,
\end{equation}
at $m=0$. The current $J_{x,\mu}$ is the conserved current associated with 
the $O(2)$ chiral symmetry \cite{Cha03}. By fitting the data for $\chi$ to 
the form given in eq. (\ref{chptchi}) we can determine $\Sigma$ accurately.

\begin{figure}[htb]
\vskip0.3in
\begin{center}
\includegraphics[width=0.45\textwidth]{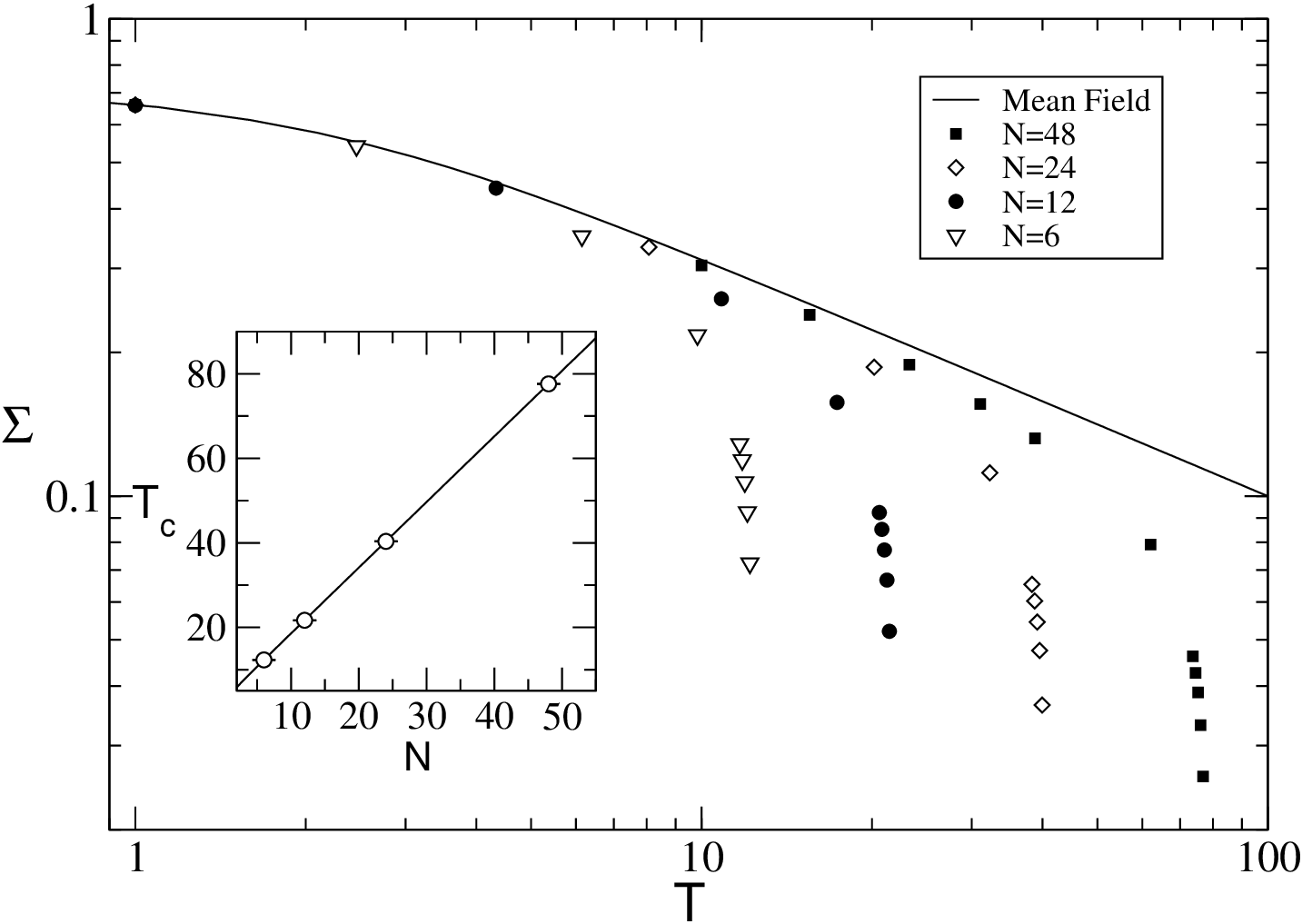}
\end{center}
\caption{\label{fig1}
Plot of $\Sigma$ vs. $T$ for various values of $N$. The errors in the 
Monte Carlo data are less than the size of the symbols. The solid line
is the mean field result given in eq. (\ref{largen}).}
\end{figure}

We have done extensive simulations for various values of $L$ and $N$
in order to extract $\Sigma$ as discussed above. In fig. (\ref{fig1}) we 
plot $\Sigma$ as a function of $T$ at $N=6,12,24,48$.
For comparison we also plot the mean field result (eq.(\ref{largen})).
As the graph shows, at a fixed value of $T$, our data approaches the
mean field prediction quite nicely as $N$ becomes large. However, for
every value of $N$, as $T$ increases the order parameter approaches 
zero at some critical temperature $T_c$. Close to $T_c$, the mean
field theory is definitely not a good approximation. Using our algorithm
we can determine $T_c$ accurately for every value of $N$ (see below).
In the inset of fig. (\ref{fig1}) we plot $T_c$ as a function of $N$. A 
fit to the form $T_c = aN + b + c/N$ yields $a=1.5525(3)$, $b=3.126(9)$ and 
$c=-0.88(4)$ with a $\chi^2/DOF = 0.73$ (solid line in the inset).

\begin{figure}[htb]
\vskip0.3in
\begin{center}
\includegraphics[width=0.45\textwidth]{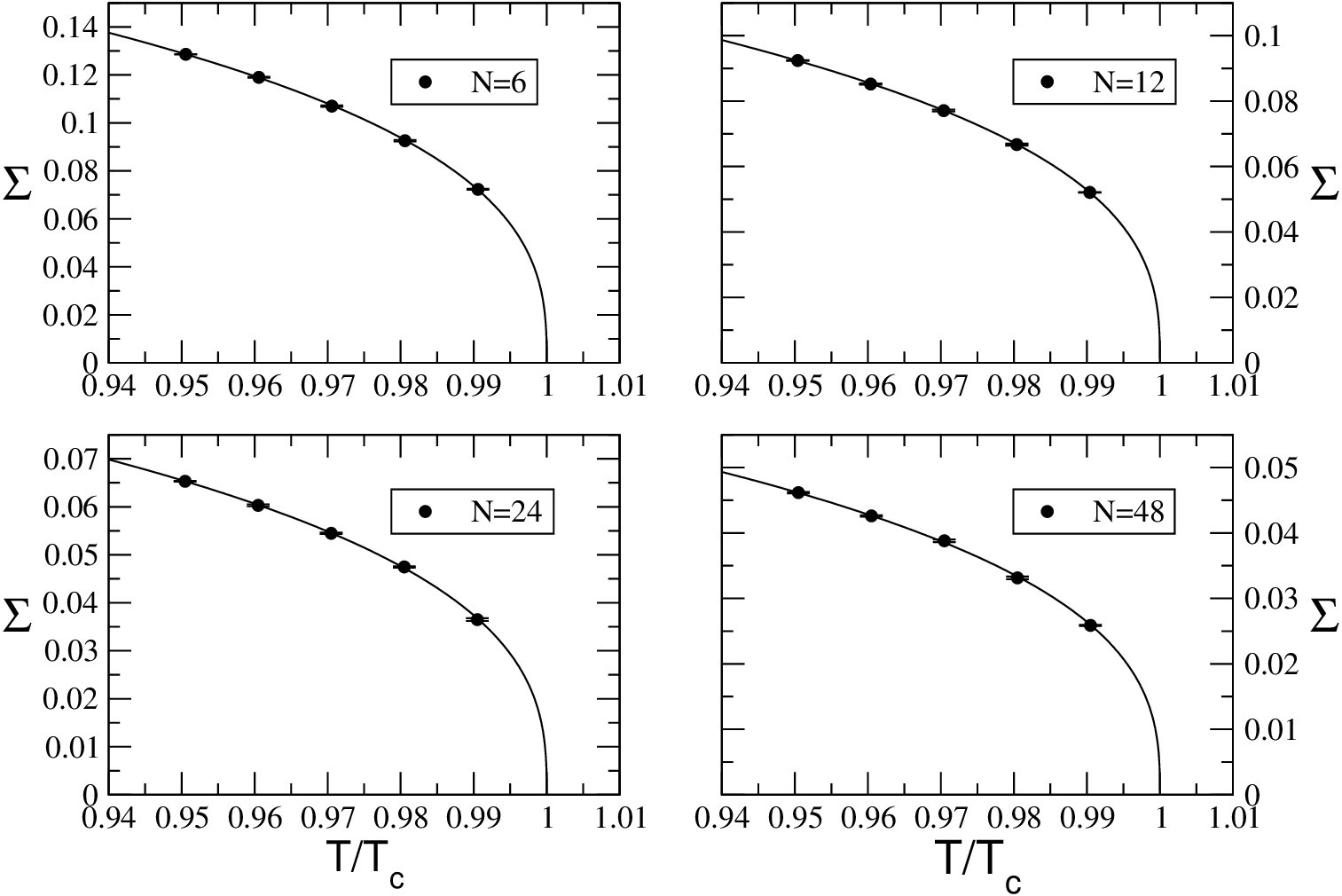}
\end{center}
\caption{\label{fig2}
Plot of $\Sigma$ vs. $T/T_c$ for various values of $N$. The solid lines
represent the fit to $A (1-T/T_c)^{0.3485}$ which is expected from
3d XY model. The values of $A$ for different $N$ are given
in the text.}
\end{figure}

One might think that it is quite easy to explain why the mean field theory 
breaks down close to $T_c$. Since $T_c$ grows as $N$, the large $N$ theory
does not know about the existence of a finite $T_c$ unless the $1/N$ 
corrections are included. We have computed these corrections and 
have found that they do not improve the situation much.
Perhaps one needs to develop a new mean field expansion where one holds 
$T/N$ fixed as $N$ becomes large. Let us refer to this as the finite $T$ 
mean field theory (FTMFT). Although we have not yet developed this mean field 
theory, we will argue that it is bound to fail close to $T_c$ since the
Ginsburg region where the 3d $XY$ universality class is observed does
not shrink with $N$. 
Indeed we find that $\Sigma \sim (1-T/T_c)^\beta$ close to $T_c$ 
where $\beta=0.3485(2)$ independent of $N$ \cite{Cam01}. A fit of the data 
to this form was used to determine $T_c$ plotted in \ref{fig1}. In 
fig. \ref{fig2} we plot $\Sigma$ as a function of $T/T_c$ for 
$0.95 \leq T/T_c \leq 1$ for $N=6,12,24,48$. The solid lines are fits to 
the form $A (1-T/T_c)^{0.3485}$. We find $A=0.3668(3), 0.2630(2), 
0.1862(2), 0.1315(2)$ for $N=6,12,24,48$ respectively all 
with a $\chi^2/DOF$ less than $1$.

\begin{figure}[htb]
\vskip0.3in
\begin{center}
\includegraphics[width=0.45\textwidth]{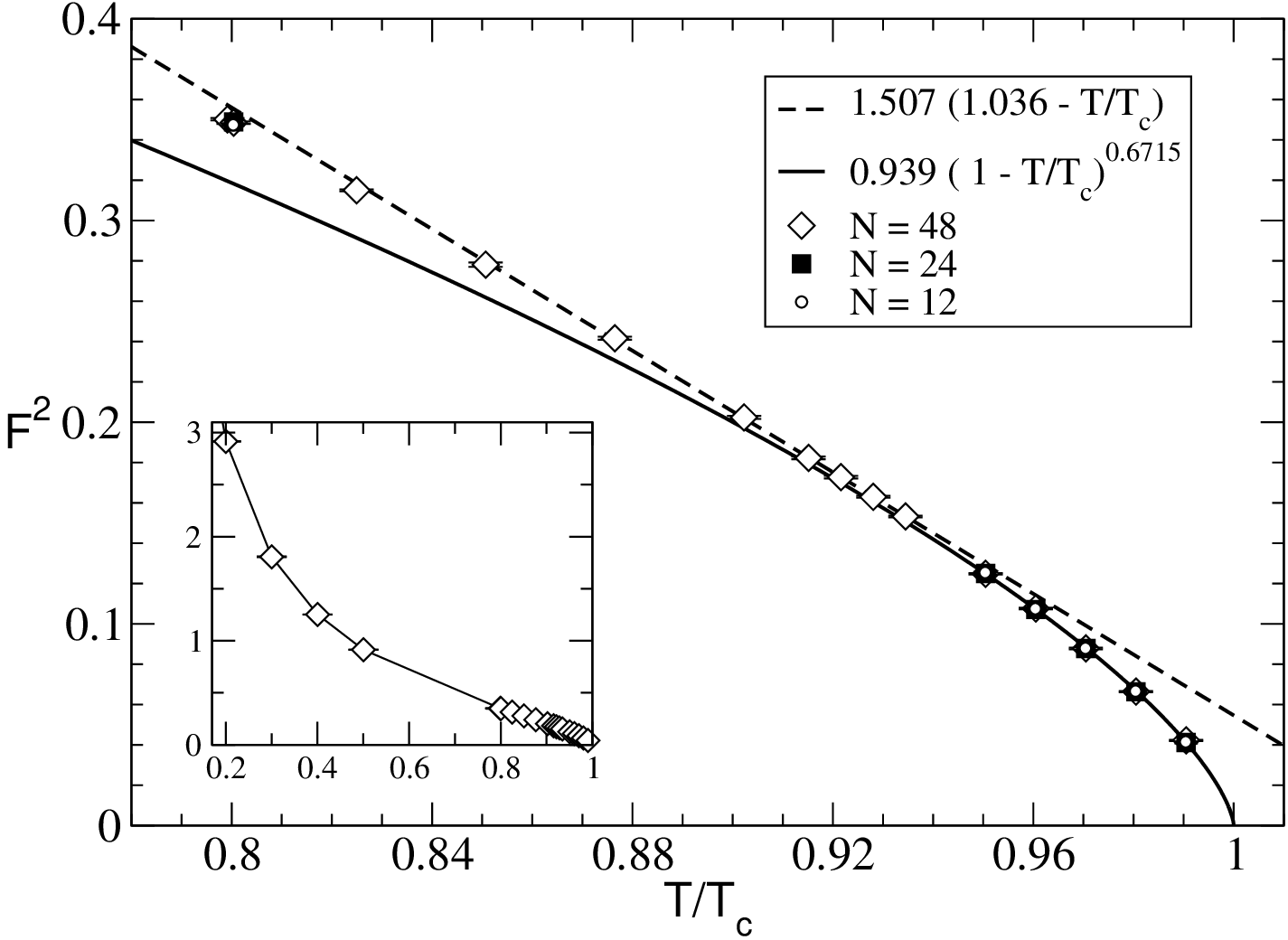}
\end{center}
\caption{\label{fig3}
Plot of $F^2$ vs. $T/T_c$ for $N=48,24,12$. The inset shows the 
same graph for a larger range of $T/T_c$ for $N=48$. The text
describes the physics of this plot.}
\end{figure}

Since $A$ decreases with increasing $N$ one might argue that
$A$ will vanish in the limit $N\rightarrow \infty$. In that case 
some higher order term will become dominant in the large $N$ limit 
and it is not possible to rule out mean field behavior. However, we do 
not think this is the case and attribute the change in $A$ to a 
renormalization effect as a function of $N$. In order to justify this, 
we next focus on a correlation length scale which does not need 
renormalization. If correlation lengths in the theory indeed scale 
with $N$, we would expect the correlation lengths to be the 
same at any fixed $T/T_c$. In our model $F^2$ can be defined as one 
such inverse correlation length scale. This implies 
$F^2 \sim B (1-T/T_c)^\nu$, where $\nu=0.6175(3)$ in the 3d $XY$ 
model \cite{Cam01}. In fig. \ref{fig3} we 
plot $F^2$ as a function of $T/T_c$ for $N=12,24,48$. When $T/T_c > 0.8$
all the points at a fixed $T/T_c$ but at different $N$ fall on top of each 
other. Further, when $0.95 \leq T/T_c \leq 1$, all the data points fit 
extremely well to the form $B (1-T/T_c)^{0.6715}$, with $B=0.939$ with 
a $\chi^2/DOF=0.65$. We believe that this is strong evidence that even 
at infinite $N$ the phase transition belongs to the 3d $XY$ universality 
class. Outside the Ginsburg region one would expect FTMFT to be valid 
\cite{Kog99}. Interestingly, when $0.85 \leq T/T_c \leq 0.93$ we find 
that the data fits to the form $B (1-T/T_c)$ with $B=1.507$ with a 
$\chi^2/DOF = 0.8$, suggesting $\nu=1$. Hence, we suspect that the 
FTMFT in our model is similar to the mean field theory in 3d $O(N)$ 
models which yields $\nu = 1$. The cross over to mean field theory
occurs when $T/T_c < 0.95$ so that the correlation length is 
$\xi \sim 1/F^2 < 8$ in lattice units.

One often hears the lore that in the large $N$ limit SCLQCD is solvable 
using mean field theory. While this is true in certain cases, in this 
article we have demonstrated that the finite temperature chiral phase 
transition belongs to the 3d $XY$ universality class even in the large $N$ 
limit. In an earlier study we found similar results in two spatial 
dimensions \cite{Cha03a}. In two dimensions continuous symmetries
cannot break at any finite temperature \cite{Mer66}. However an $O(2)$ 
symmetry is special and a phase transition in the 
Berezinski-Kosterlitz-Thouless (BKT) universality class is 
possible \cite{Ber71}. The BKT prediction 
for our model is that $\chi \sim L^{2-\eta}$ where $\eta$ is a function 
of temperature $0 \leq \eta(T) \leq 1/4$ when $T \leq T_c$. At the critical 
temperature $\eta=1/4$. We found all this to be true even at large $N$.
Interestingly, according to Witten a large $N$ mean field theory 
would find $\eta \sim 1/N$ \cite{Wit78}. While our results agree
with this observation at a fixed temperature $T$, we find that
if $T/T_c$ is held fixed then $\eta$ approaches a non-zero value 
in the predicted range showing that the mean field approach again 
breaks down in the large $N$ theory close to the phase transition.

Our study shows that the large $N$ limit may not always be able to suppress
infrared fluctuations close to second order phase transitions. In a
sense this is an indication that the perturbation expansion starting 
from a mean field solution breaks down. In other words a careful analysis 
of the FTMFT should be able to reveal this. This has not yet been done 
and is a useful project for the future. It can help us classify the types 
of large $N$ models where mean field theory can break down. 

\section*{Acknowledgments}

We thank K. Splittorff, M. Stephanov and D. Toublan for helpful 
discussions. This work was supported in part by grants DE-FG-96ER40945 
and DE-FG02-03ER41241 from the Department of Energy (DOE). SC would also 
like to thank the Institute of Nuclear Theory for hospitality where part 
of this work was done. The computations were performed on CHAMP, a 
computer cluster funded in part by the DOE and computers of 
Dr. R. Brown who graciously allowed us to use them when they were 
available.


\begin{thebibliography}{99}

\bibitem{Zin}
J.~Zinn-Justin, {\em Quantum Field Theory and Critical Phenomena},
4th edition, Oxford University Press.

\bibitem{Amit}
D.~J.~Amit, {\em Field Theory, the renormalization group and
Critical Phenomena}, Revised 2nd Edition, World Scientific.

\bibitem{Mos03}
M. Moshe and J. Zinn-Justin, Phys. Rep. 385, 69 (2003).

\bibitem{Koc95}
A. Kocic and J. Kogut, Phys. Rev. Lett. 74, 3109 (1995).

\bibitem{Kog99}
J.~Kogut, M.~Stephanov and C.~G.~Strouthos, Phys.Rev.D58, 096001 (1998).

\bibitem{Ste04}
M. Stephanov, Private Communication.

\bibitem{Cos04}
C.~G.~Strouthos, arXiv:hep-lat/0410003.

\bibitem{Hei72}
O.~J.~Heilmann and E.~H.~Lieb, Commun. Math. Phys. 25, 190 (1972).

\bibitem{Kas63}
P. W. Kasteleyn, Physica 27, 1209 (1961); J. Math. Phys. 4, 287 (1963).

\bibitem{Fis63}
M. Fisher and J. Stephenson, Phys. Rev. 132, 1411 (1963);

\bibitem{Rok88}
D. S. Rokhsar and S. A. Kivelson, Phys. Rev. Lett. 61, 2376 (1998).

\bibitem{And87}
P. W. Anderson, Science 235, 1196 (1987).

\bibitem{Moe01}
R. Moessner and S.L. Sondhi, Phys. Rev. Lett 86, 1881 (2001).

\bibitem{Moe02}
R. Moessner, S.L. Sondhi and E. Fradkin, Phys. Rev. B65, 024504 (2002).

\bibitem{Boy92}
G.~Boyd, J.~Fingberg, F.~Karsch, L.~Karkkainen and B.~Petersson 
Nucl. Phys. B376, 199 (1992).

\bibitem{Klu81}
H. Kluberg-Stern, A. Morel, O. Napoly and B. Petersson,
Nucl. Phys. B190, 504 (1981).

\bibitem{Ros84}
P.~Rossi and U.~Wolff, Nucl. Phys. B248, 105 (1984).

\bibitem{Ada03}
D.H.~Adams and S.~Chandrasekharan, Nucl. Phys. B662, 220 (2003).

\bibitem{Cha03}
S.~Chandrasekharan and F.J.~Jiang, Phys. Rev. D68, 091501(R) (2003).

\bibitem{Bil92}
N.Bili\'{c}, F.Karsch and K.Redlich, Phys. Rev. D45, 3228 (1992).

\bibitem{Has90}
P.~Hasenfratz and H.~Leutwyler, Nucl. Phys. B343, 241 (1990).

\bibitem{Cam01}
M.~Campostrini, M.~Hasenbusch, A.~Pelissetto, P.~Rossi and E.~Vicari,
Phys. Rev. B63, 214503 (2001).

\bibitem{Cha03a}
S.~Chandrasekharan and C.~G.~Strouthos, Phys. Rev. D68, 091502(R) (2003).

\bibitem{Mer66}
N.~D.~Mermin and H.~Wagner, Phys. Rev. Lett. 17, 1133 (1966);
S.~Coleman, Commun. Math. Phys. 31, 259 (1973).

\bibitem{Ber71}
V.L.~Berezinski, Sov. Phys. JETP 34, 610 (1971);
J.M.~Kosterlitz and D.J.~Thouless, J.Phys. C 6, 1181 (1973).

\bibitem{Wit78}
E. Witten, Nucl. Phys. B145, 110 (1978).

\end{thebibliography}
\end{document}